\documentclass{amsart}%
\usepackage{amssymb}
\usepackage{amsfonts}
\usepackage{amsmath}
\usepackage{graphicx}%
\theoremstyle{plain}

\newtheorem{lemma}{Lemma}

\numberwithin{equation}{section}
\begin{document}
\title[Quantum Densities, Distances and Fidelities]{Contravariant Densities, Complete Distances and Relative Fidelities for
Quantum Channels }
\author{Viacheslav Belavkin}
\address{School of Mathematics, University of Nottingham, NG7 2RD, UK}
\email{vpb@maths.nott.ac.uk}
\urladdr{http://www.maths.nott.ac.uk/personal/vpb/}
\thanks{}
\thanks{Author acknowledge EEC support through the ATESIT project IST-2000-29681 and
QP\&Applications RTN2-2001-00378 grant.}
\date{October 15, 2003}
\subjclass{ }
\dedicatory{In celebration of the 100th anniversary of the birth of John von Neumann }
\begin{abstract}
Introducing contravariant trace-densities for quantum states, we restore
one-to-one correspondence between quantum operations described by normal CP
maps and their trace densities as Hermitian-positive operator-valued
contravariant kernels. The CB-norm distance between two quantum operations is
explicitly expressed in terms of these densities as the supremum over the
input states. A larger C-distance is given as the natural norm-distance for
the channel densities, and another, Helinger type complete distance
(CH-distance), related to the minimax mean square fidelity optimization
problem by purification of quantum channels, is also introduced and evaluated
in terms of their contravariant trace-densities. It is proved that the CH
distance between two channels is equivalent to the CB distance. An operational
meaning for these distances and relative complete fidelity for quantum
channels is given in terms of quantum encodings producing optimal
entanglements of quantum states for an opposite and output systems.

\end{abstract}
\maketitle

\section{Introduction}

Quantum channels, which are usually described by trace preserving operations
$\mathrm{T}:\mathcal{A}_{\ast}\rightarrow\mathcal{B}_{\ast}$, preadjoint to
normal completely positive (CP) maps $\Psi=\mathrm{T}^{\ast}$ of the output
algebra $\mathcal{B}=\mathcal{B}\left(  \mathcal{H}\right)  $ into an input
operator algebra $\mathcal{A}$ with respect to the Hilbert-Schmidt pairing%
\[
\left\langle \varrho|\Psi\left(  B\right)  \right\rangle :=\tau\left[
\varrho^{\dagger}\Psi\left(  B\right)  \right]  =\tau\left[  \mathrm{T}\left(
\varrho\right)  ^{\dagger}A\right]  \equiv\left\langle \mathrm{T}\left(
\varrho\right)  |B\right\rangle ,
\]
can be completely characterized in terms of their positive densities
$\Psi_{\tau}$ with respect to the standard trace $\tau=\mathrm{tr}$ on
$\mathcal{B}$, see for example \cite{BeOh02}. Such densities, defined as
non-commutative Radon-Nikodym derivatives \cite{BeSt86} of $\Psi$ with respect
to $\tau$, even in the case of states $\Psi=\rho$ are actually different from
the usually used densities $\varrho$. The Radon-Nikodym derivative of a normal
state $\rho$ on the matrix algebra $\mathcal{A}=\mathcal{B}\left(
\mathcal{H}\right)  $ with respect to the reference trace $\tau=\mathrm{tr}$
is canonically identified by a normal representation with not the covariant
density matrix $\varrho\in\mathcal{A}_{\ast}$ but with the contravariant
density $\rho_{\tau}$ as the complex conjugated $\rho_{\tau}=\bar{\varrho}$
(or equivalently transposed, $\rho_{\tau}=\tilde{\varrho}$) matrix in the
commutant of the standard representation of $\mathcal{A}$. In general the
contravariant state densities $\rho_{\tau}$ are uniquely defined as affiliated
elements of the opposite algebra $\overline{\mathcal{A}}$, the positive
normalized elements of which describe normal states $\rho$ on $\mathcal{A}$
majorized by $\tau$ as $\rho\left(  A\right)  =\left\langle A,\rho_{\tau
}\right\rangle $ with respect to the bilinear pairing $\left\langle
A,\rho_{\tau}\right\rangle =\left\langle \bar{\rho}|A\right\rangle $ of
$\mathcal{A}$ and $\overline{\mathcal{A}}_{\ast}$. They are more suitable for
operational generalizations than the usual, covariant densities $\rho_{\ast
}=\varrho$ which retain only partial (transpose) positivity when they are
replaced by covariant channel densities $\Psi_{\ast}\in\mathcal{A}%
\otimes\mathcal{B}_{\ast}$ defining the maps $\Psi$ by the partial trace
$\Psi\left(  B\right)  =\mathrm{tr}_{\mathcal{B}}\left(  I\otimes B\right)
\Psi_{\ast}$. The positive, contravariant densities $\Psi_{\tau}$ describe CP
maps $\Psi$ by partial tracing corresponding to $\left\langle B,\sigma_{\tau
}\right\rangle =\mathrm{tr}\widetilde{B}\sigma_{\tau}$, and they transform the
contravariant input densities $\rho_{\tau}$ into contravariant densities
$\sigma_{\tau}=\bar{\varsigma}$ of output states $\varsigma=\mathrm{T}\left(
\varrho\right)  $ by partial tracing%
\[
\left\langle \Psi_{\tau},\rho_{\tau}\right\rangle =\mathrm{tr}_{\mathcal{A}%
}\Psi_{\tau}\left(  \tilde{\rho}_{\tau}\otimes I_{\mathcal{B}}\right)  .
\]

In the case of infinite dimensional Hilbert space $\mathcal{H}$, the channel
densities $\Psi_{\tau}$, unlike the state densities, might be unbounded even
in the simple cse $\mathcal{B}=\mathcal{B}\left(  \mathcal{H}\right)  $, and
in general should be understood in a distribution sense. Such densities were
defined in \cite{BeSt86} as Hermitian-positive kernels, characterized as
unbounded self-adjoint operators in Stinespring representation \cite{Sti55} of
another, dominating CP map $\Phi$. In the case of tracial $\Phi\left(
B\right)  =I_{\mathcal{A}}\mathrm{tr}B$ the channel densities $\Psi_{\tau}$
are defined as Hermitian-positive kernels affiliated to the tensor product
$\mathcal{A}\otimes\overline{\mathcal{B}}$ of the input channel algebra
$\mathcal{A}$ with the opposite (transposed) algebra $\overline{\mathcal{B}}$
to the output $\mathcal{B}$. The Schmidt decompositions of the positive
densities $\Psi_{\tau}$ are in one-to-one correspondence with the Kraus
decompositions of the channels $\Psi$, and the spectral decompositions of the
positive self-adjoint operators $\Psi_{\tau}$, corresponding to the orthogonal
Kraus decompositions of $\Psi$, completely describe the properties of quantum
channels in terms of their spectral measures on $\mathbb{R}_{+}$.

Here we develop a consistent metric space theory of contravariant quantum
channel kernels describing quantum operations in terms of trace-pairings of
the observables and the contravariant densities respectively to generalized
traces, introduced in \cite{BeOh02}. We discuss several distances for
comparing two quantum channels $\Phi$ and $\Psi$, among them the complete
boundedness (CB) distance and the complete Helinger (CH), or operational Bures
distance, and define the complete relative fidelity of these channels. All
these distances, conditioned upon the input state, are explicitly evaluated in
terms of the contravariant densities $\Phi_{\tau},\Psi_{\tau}$\ of the
channels. Thus, the CB distance is expressed as the maximum $D_{\mathrm{cb}%
}(\Phi,\Psi)=\sup_{\rho}D_{\mathrm{cb}}^{\rho}(\Phi,\Psi)$ of the standard
entanglement trace distance
\[
D_{\mathrm{cb}}^{\rho}(\Phi,\Psi)=\operatorname{Tr}\left\vert \left(
\rho_{\ast}^{1/2}\otimes I_{\mathcal{B}}\right)  \left(  \Phi_{\tau}%
-\Psi_{\tau}\right)  \left(  \rho_{\ast}{}^{1/2}\otimes I_{\mathcal{B}%
}\right)  \right\vert ,
\]
over the input state densities $\rho_{\ast}\in\mathcal{A}_{\ast}$ as
contravariant densities of normal states on $\overline{\mathcal{A}}$. The CH
distance $d_{\mathrm{c}}\left(  \Phi,\Psi\right)  $ between two quantum
channels $\Phi$ and $\Psi$ is found as $d_{\mathrm{c}}\left(  \Phi
,\Psi\right)  ^{2}/2=1-f_{\mathrm{c}}\left(  \Phi,\Psi\right)  $ in terms of
the minimum $f_{\mathrm{c}}\left(  \Phi,\Psi\right)  =\inf_{\rho}%
f_{\mathrm{c}}^{\rho}\left(  \Phi,\Psi\right)  $ of the standard entanglement
relative fidelity
\[
f_{\mathrm{c}}^{\rho}\left(  \Phi,\Psi\right)  =\operatorname{Tr}\left\vert
\Phi_{\tau}^{1/2}\left(  \rho_{\ast}\otimes I_{\mathfrak{h}}\right)
\Psi_{\tau}^{1/2}\right\vert
\]
for a given $\rho$. Here and below $\left\vert A\right\vert $ means the
"modulus" of an operator $A$, for which under the trace can be taken any of
the expressions $\sqrt{A^{\dagger}A}$ and $\sqrt{AA^{\dagger}}$ (which are
different in the case of not normal $A$).

CB distance has recently found an extensive use quantum information theory,
see for example the review paper [\cite{KrWe04}]. While the CH distance, as
will be shown here, is topologically equivalent to the CB distance between two
channels, from the operational and computational point of view this new
(Helinger) complete distance have certain advantages over the CB distance.

Note that in the finite dimensional case one can use for discriminating of the
channels the standard entanglement relative fidelity with respect to the input
tracial state $\rho$ corresponding to $\rho_{\ast}=d^{-1}I$. In fact such
fidelity has been recently suggested by Raginsky \cite{Rag01} who used to
define his fidelity relative to the maximal entangled state. However the
corresponding fidelity distance is not equivalent to the CB distance, and
there is no such measure in the infinite dimensional case. The normalized
tracial states do not exist on type one algebras if $d=\infty$, and there is
no maximal entangled state. We prove that our complete fidelity distance is
equivalent to the CB distance, and give an operational interpretation of this
fidelity in terms of a minimax problem for quantum encodings and
decompositions, purifying the channels, in complete parallel to Uhlman's
theorem \cite{Uhl76}.

\section{Some facts and notation}

\emph{Quantum state} $\sigma$, identified as usual with a list of all
expectations $\sigma\left(  A\right)  $, is defined as a linear normalized
functional $A\mapsto$ $\sigma\left(  A\right)  $ on the algebra $\mathcal{B}%
\left(  \mathcal{H}\right)  $ of all bounded operators $A$ in a separable
Hilbert space $\mathcal{H}$. The functional also satisfies the positivity
condition $\sigma\left(  A^{\dagger}A\right)  \geq0$ for any $A\in
\mathcal{B}\left(  \mathcal{H}\right)  $ such that $\sigma\left(  I\right)
=1$ for the identity operator $I$ on $\mathcal{H}$. As usual we shall consider
only normal states $\sigma\left(  A\right)  =\left\langle A,\sigma
\right\rangle $, identifying them with \emph{density operators} $\sigma
=\sum|i\rangle\sigma^{ik}\langle k|$ defining the expectations $\sigma\left(
A\right)  $ as sums, or the series%
\begin{equation}
\left\langle A,\sigma\right\rangle =\sum_{ik}a_{ik}\sigma^{ik}\equiv
a_{ik}\sigma^{ik},\;\;\;a_{ik}=\left\langle i|A|k\right\rangle .
\label{pairing}%
\end{equation}
Here $\sigma$ is given in an orthogonal basis of real units $\langle
i|\in\mathcal{H}$ and $|i\rangle=\langle i|^{\dagger}$ as a Hermit-positive
$\sigma\geq0$ matrix of $\sigma^{ik}=\sigma\left(  |i\rangle\langle k|\right)
$ with the unit trace%
\[
\operatorname{Tr}\sigma=\sum_{i}\sigma^{ii}\equiv\left\langle I,\sigma
\right\rangle =1.
\]

For the reason which will be explained later we prefer to use the tensor form
(\ref{pairing}) for the bilinear pairing $A$ and $\sigma$ rather than
sesquilinear form $\operatorname{Tr}\bar{\sigma}A=\sum\sigma_{ik}a_{ki}$ which
pairs $A$ with the covariant form $\bar{\sigma}=\left[  \sigma_{ik}\right]  $
of the density matrix $\sigma=\left[  \sigma^{ik}\right]  $ in terms of the
complex conjugated elements $\bar{\sigma}^{ik}\equiv\sigma_{ik}=\sigma^{ki}$,
coinciding with the transposed ones due to $\sigma^{\dagger}=\sigma$.
Equivalently (\ref{pairing}) can be written as $\left\langle A,\sigma
\right\rangle =\operatorname{Tr}\tilde{\sigma}A$, where $\tilde{\sigma}$
denotes the transposed matrix $\left[  \sigma^{ki}\right]  $. The matrix
$\left[  \sigma^{ik}\right]  $ is called the \emph{contravariant density} of
the state $\sigma$ with respect to the standard trace $\tau=\operatorname{Tr}%
$. As we shall see, contravariant density gives more adequate representation
for quantum channels than the matrix $\bar{\sigma}=\left[  \left\langle
i|\bar{\sigma}|k\right\rangle \right]  =\tilde{\sigma}$, and its name can be
explained in terms of the \emph{contravariant transformations}. If~$\left\{
v_{i}\right\}  $ is an orthogonal basis of $\mathcal{H}$ in terms of rows
$v_{i}$ such that their Hermitian adjoints $v_{i}^{\dagger}$ form a unitary
matrix $U=\left(  v_{i}^{\dagger}\right)  $, then the quantum expectations
$\sigma\left(  A\right)  $ do not change if the corresponding densities
$\sigma\in\mathcal{B}_{\intercal}\left(  \mathcal{H}\right)  $ are
contravariantly transformed as $U\sigma U^{\dagger}=v_{i}^{\dagger}v_{k}%
\sigma^{ik}$ with respect to the observables transformation $V^{\dagger
}AV=a_{ik}v^{i\dagger}v^{k}$, where $V=\left[  v^{k}\right]  $ is given as a
column of the complex conjugate rows $v^{i}=\bar{v}_{i}$ such that
$V_{\intercal}:=U=\widetilde{V}$. (They are inversely transformed in the case
$\overline{U}=U^{\dagger}$ of the \emph{symmetric transformations}
$\widetilde{V}=V$, in which case $U=V$.)

Recall that the linear span of the convex set $\mathcal{S}_{\tau}%
=\mathcal{S}\left(  \mathcal{H}\right)  $ of all trace-densities $\sigma$ for
normal states on $\mathcal{B}\left(  \mathcal{H}\right)  $ is a Banach space
$\mathcal{B}_{\intercal}\left(  \mathcal{H}\right)  $ with respect to the
trace norm $\Vert A\Vert_{\intercal}:=\operatorname{Tr}\sqrt{A^{\dagger}A}$,
consisting of all trace class operators in $\mathcal{H}$, and its dual
$\mathcal{B}_{\intercal}\left(  \mathcal{H}\right)  ^{\intercal}$, the space
of all bounded functionals, coincides with $\mathcal{B}\left(  \mathcal{H}%
\right)  $ with respect to the pairing (\ref{pairing}). Given a $\sigma
\in\mathcal{B}_{\intercal}\left(  \mathcal{H}\right)  $, one can define the
pair-transposed element $A^{\intercal}$ as the functional%
\[
A^{\intercal}\left(  \sigma\right)  =\left\langle A,\sigma\right\rangle .
\]
on the \emph{predual} space $\mathcal{B}_{\intercal}\left(  \mathcal{H}%
\right)  $ given by an operator $A\in\mathcal{B}\left(  \mathcal{H}\right)  $.
Note that $A^{\intercal}$ is not the transposed operator in $\mathcal{H}$ but
is described in terms of the transposed operator $\widetilde{A}=\overline
{A}^{\dagger}$ as $A^{\intercal}\left(  \sigma\right)  =\operatorname{Tr}%
\widetilde{A}\sigma$. The transposition $A\mapsto\widetilde{A}$ is related not
to the operator but to the Hilbert space pairing $\left\langle \bar{\varphi
},\psi\right\rangle =\left\langle \varphi|\psi\right\rangle $ defining the
\emph{vector transpose }$\psi\mapsto\tilde{\psi}$ as the reverse to
$|\psi\rangle\mapsto\langle\bar{\psi}|$ by identifying the complex conjugate
elements $\bar{\psi}\in\mathcal{H}$ with bra-vectors $\langle\psi|$.

All of that can be easily generalized \cite{BeOh02} to an arbitrary operator
subalgebra $\mathcal{B}\subseteq\mathcal{B}\left(  \mathcal{H}\right)  $ with
the standard, or a nonstandard normal faithful semifinite trace \cite{Sim79}
$\nu$ when the standard one, $\tau=\operatorname{Tr}$, is not semifinite, e.g.
is trivial on $\mathcal{B}$ in the sense that $\operatorname{Tr}A^{\dagger
}A=\infty$ for all operators $A\neq0$ from the algebra $\mathcal{B}$. Such a
trace $\nu$ is defined as a nonnormalized state on $\mathcal{B}$, finite on a
weakly dense part of $\mathcal{B}$, with the property $\nu\left(  A^{\dagger
}A\right)  =\nu\left(  AA^{\dagger}\right)  $ and separating $\mathcal{B}$ in
the sense that $\nu\left(  A^{\dagger}A\right)  =0\Rightarrow A=0$. A quantum
system will be called \emph{semifinite} if it is described by the semifinite
algebra, i.e. admits a semifinite, not necessarily standard, trace $\nu$.
(There exist also quantum infinite systems which are not semifinite.) The only
difference is that the predual space $\mathcal{B}_{\intercal}$ of
contravariant densities $\sigma$ with respect to the trace $\nu$ on
$\mathcal{B}$ may not be apart of the algebra $\mathcal{B}$ but a part of an
\emph{opposite} algebra $\overline{\mathcal{B}}$, or, if unbounded, affiliated
to $\overline{\mathcal{B}}$. The opposite algebra can be defined on the same
Hilbert space equipped with a complex conjugation $\chi\mapsto\bar{\chi}$,
i.e. isometric involution in $\mathcal{H}$, as the subalgebra $\overline
{\mathcal{B}}=\left\{  A\in\mathcal{B}\left(  \mathcal{H}\right)
:\overline{A}\in\mathcal{B}\right\}  $ of complex conjugated operators
$\overline{A}=\widetilde{A}^{\dagger}$. This $\overline{\mathcal{B}}$,
equipped with the \emph{reference} trace $\bar{\nu}\left(  \overline
{A}\right)  :=\overline{\nu\left(  A\right)  }$, does not necessarily coincide
with $\left(  \mathcal{B},\nu\right)  $ as in the simple case $\mathcal{B}%
=\mathcal{B}\left(  \mathcal{H}\right)  $. The "opposite" trace $\mu=\bar{\nu
}$, coinciding with $\tilde{\nu}\left(  A\right)  =\nu\left(  \widetilde
{A}\right)  $ on the opposite algebra $\overline{\mathcal{B}}$, defines the
$\mu$-pairing%
\begin{equation}
\left\langle A,\sigma\right\rangle _{\mu}=\mu\left(  \widetilde{A}%
\sigma\right)  ,\;\;A\in\mathcal{B},\sigma\in\mathcal{B}_{\intercal},
\label{mupairing}%
\end{equation}
generalizing (\ref{pairing}). Note that in the symmetric case $\mathcal{B}%
=\overline{\mathcal{B}}$ $\ $the trace $\mu$ is not distinguished from $\nu$:
$\bar{\nu}\left(  A\right)  =\nu\left(  \widetilde{A}\right)  =\nu\left(
A\right)  $.

Below $\mathcal{S}_{\tau}\subseteq\mathcal{A}_{\ast}$ denotes the convex set
of all positive normalized contravariant densities $\varrho\geq0$,
$\tau\left(  \varrho\right)  =1$ for normal states on the von Neumann algebra
$\overline{\mathcal{A}}$ with respect to a normal faithful semifinite trace
$\lambda=\bar{\tau}$. They can also be regarded as covariant densities,
$\rho_{\ast}=\varrho\in\mathcal{A}_{\ast}$ of the normal states $\rho\left(
A\right)  =\tau\left(  \varrho A\right)  $ on the opposite algebra
$\mathcal{A}$ with the reference trace $\tau$.

\section{Quantum Bures and trace distances}

The difference between two states $\rho,\sigma\in\mathcal{S}\left(
\mathcal{H}\right)  $ is usually measured by \emph{trace-norm distance}
\begin{equation}
D(\rho,\sigma):=\operatorname{Tr}\left\vert \rho-\sigma\right\vert \equiv
\Vert\rho-\sigma\Vert_{\intercal}, \label{eq:tnd}%
\end{equation}
given in terms of the modulus-difference of their trace densities $\rho
=\rho_{\tau}$ and $\sigma=\sigma_{\tau}$. It is also defined for non positive
$\rho$ and $\sigma$ and thus doesn't have much information-operational
meaning. Another measure of this difference is relative entropy \cite{OhPe93}
which has clearly more information-operational meaning. However the quantum
relative entropy is not symmetric and not unique, and is not a distance on
$\mathcal{S}_{\tau}$.

The natural operational distance between two quantum states is \emph{quantum
Bures distance} which can be defined as the square root of the minimal
Euclidean squared distance%
\begin{equation}
d\left(  \rho,\sigma\right)  ^{2}:=\inf_{\chi^{\dagger}\chi=\rho,\psi
^{\dagger}\psi=\sigma}\operatorname{Tr}\left(  \chi-\psi\right)  ^{\dagger
}\left(  \chi-\psi\right)  . \label{statehdist}%
\end{equation}
Here the infimum is taken over all all Schmidt decompositions of the
trace-densities $\rho$ and $\sigma$. Uhlman's theorem \cite{Uhl76} states that
this infimum is actually achieved at the value $d\left(  \rho,\sigma\right)
^{2}/2=1-f\left(  \rho,\sigma\right)  $, where
\begin{equation}
f\left(  \rho,\sigma\right)  =\operatorname{Tr}\sqrt{\rho^{1/2}\sigma
\rho^{1/2}}=:\operatorname{Tr}\left\vert \rho^{1/2}\sigma^{1/2}\right\vert
:=\operatorname{Tr}\sqrt{\sigma^{1/2}\rho\sigma^{1/2}} \label{eq:statefid}%
\end{equation}
is the relative fidelity of the states $\rho$ and $\sigma$. Note that this can
obviously be written as the trace of the matrix%
\[
\rho^{1/2}\sqrt{\rho^{1/2}\sigma\rho^{1/2}}\rho^{-1/2}\equiv\sqrt{\rho\sigma
}\equiv\sigma^{-1/2}\sqrt{\sigma^{1/2}\rho\sigma^{1/2}}\sigma^{1/2}.
\]
giving a much simpler formula $\operatorname{Tr}\sqrt{\rho\sigma}$ for the
fidelity $f\left(  \rho,\sigma\right)  $. It is also valid even if $\rho$
and/or $\sigma$ are not invertible as soon as the product $\rho\sigma=T\Lambda
T^{-1}$ remains similar to a positive diagonal matrix $\Lambda$ to make sense
of $\sqrt{\rho\sigma}=T\Lambda^{1/2}T^{-1}$.

Thus, quantum Bures distance is a noncommutative generalization of the
statistical Helinger distance which in the case of the commutative $\rho$ and
$\sigma$ has the familiar form $\operatorname{Tr}\left(  \sqrt{\rho}%
-\sqrt{\sigma}\right)  ^{2}$. Since Schmidt decompositions $\psi^{\dagger}%
\psi=\sum\psi^{\dagger}|j\rangle\langle j|\psi$ correspond to purifications of
the states $\sigma$, the infimum (\ref{statehdist}) has clear
information-operational meaning.

Although the squared fidelity distance $d^{2}$ is smaller \cite{Uhl76} than
$D$, they achieve the same maximal value $d^{2}=2=D$ on $\mathcal{S}\left(
\mathcal{H}\right)  $ when $f=0$, i.e. when the range of $\rho$ is orthogonal
to the range of $\sigma$. In fact, as it follows from the inequality
$D\leq2\sqrt{1-f^{2}}$, see for example \cite{NiCh00}, $D$ cannot be larger
than $2d$. Thus both distances are topologically equivalent,
\begin{equation}
d\left(  \rho,\sigma\right)  ^{2}\leq D\left(  \rho,\sigma\right)
\leq2d\left(  \rho,\sigma\right)  , \label{eq:fvd}%
\end{equation}
and $d=0=D$ iff $f=1$, i.e. $\rho=\sigma$.

All of that can be easily generalized to a more general semifinite algebra
$\mathcal{B}\subset\mathcal{B}\left(  \mathcal{H}\right)  $ with respect to a
trace $\nu$, inducing the (nonstandard) reference trace $\mu=\bar{\nu}$ on the
opposite algebra $\overline{\mathcal{B}}$, and the pairing of $\mathcal{B}$
with the predual space $\mathcal{B}_{\mathcal{\intercal}}$ affiliated to the
opposite algebra $\overline{\mathcal{B}}$ is understood in the sense of
(\ref{mupairing}). The only difference is that the contravariant densities
$\rho=\rho_{\mu}$, $\sigma=\sigma_{\mu}$, normalized with respect to trace
$\mu=\bar{\nu}$ on $\overline{\mathcal{B}}$, may not be bounded, but they are
still uniquely described by the positive selfadjoint operators in
$\mathcal{H}$ affiliated to $\overline{\mathcal{B}}$. The trace and the Bures
distance formulae remain the same with the reference trace $\mu=\bar{\nu}$
replacing the standard trace, and the fidelity formula with respect to the
arbitrary trace $\mu$ is generalized to%
\begin{equation}
f\left(  \rho,\sigma\right)  =\mu\left(  \sqrt{\rho\sigma}\right)  =\mu\left(
\left\vert \rho^{1/2}\sigma^{1/2}\right\vert \right)  =\mu\left(  \sqrt
{\sigma\rho}\right)  . \label{mufid}%
\end{equation}
Here $\sqrt{\rho\sigma}$ (and $\sqrt{\sigma\rho}$) is still well-defined as
the function $T\sqrt{\Lambda}T^{-1}$ of the operator $\rho\sigma=T\Lambda
T^{-1}$ (and $\sqrt{\sigma\rho}=T\Lambda T^{-1}$) as being similar to a
positive diagonal operator $\Lambda$ with respect to a similarity
transformation $T$. If $\rho$ and $\sigma$ are invertible, one can take either
$\sigma^{-1/2}U$ or $\rho^{1/2}V^{\dagger}$as such $T$ (respectively either
$\rho^{-1/2}V$ or $\sigma^{1/2}U^{\dagger}$), where $U$ and $V$ are unitary
transformation diagonalizing respectively $\sigma^{1/2}\rho\sigma^{1/2}$ and
$\rho^{1/2}\sigma\rho^{1/2}$.

\section{Operational densities and quantum channels}

In order to compare \emph{quantum channels} we need to generalize these
results to linear completely positive (CP) trace preserving mappings
$\Phi_{\intercal}$ from the predual space $\mathcal{A}_{\intercal}$ of the
"Alice" algebra $\mathcal{A}\subseteq\mathcal{B}\left(  \mathfrak{g}\right)  $
on an \emph{input} Hilbert space $\mathfrak{g}$ into the predual space
$\mathcal{B}_{\intercal}$ of the "Bob" algebra $\mathcal{B}\subseteq
\mathcal{B}\left(  \mathfrak{h}\right)  $ on the same or different
\emph{output} Hilbert space $\mathfrak{h}$. The notation $\mathcal{H}$ will be
kept for the Hilbert product $\mathfrak{g}\otimes\mathfrak{h}$ with the total
trace $\operatorname{Tr}=\tau_{\mathfrak{g}}\otimes\tau_{\mathfrak{h}}$
inducing the trace $\lambda\otimes\bar{\mu}$ on the entangled input-output
system $\overline{\mathcal{A}}\otimes\mathcal{B}$ as the product of the
standard traces $\lambda=\tau_{\mathfrak{g}}|\overline{\mathcal{A}}$ and
$\bar{\mu}=\tau_{\mathfrak{h}}|\mathcal{B}$ if $\tau_{\mathfrak{g}%
}=\mathrm{tr}_{\mathfrak{g}}$ and $\tau=\mathrm{tr}$ remain on these
subalgebras semifinite. Otherwise we should take non-standard reference traces
$\tau$ on $\mathcal{A}$ and $\nu$ on $\mathcal{B}$ and define $\lambda
\otimes\bar{\mu}$ as the opposite trace to $\tau\otimes\mu$ on
$\mathcal{A\otimes}\overline{\mathcal{B}}$.

As in the case of the states it is more convenient to describe quantum
channels by the normal unital CP operations defined as the maps $\Phi
:\mathcal{B}\rightarrow\mathcal{A}$ on the \emph{output algebra} $\mathcal{B}$
into the \emph{input algebra} $\mathcal{A}$, like the expectations $\sigma$
mapping $\mathcal{B}$ into the trivial algebra $\mathcal{A}=\mathbb{C}$. These
can always be defined as the dual $\Phi_{\intercal}^{\intercal}$ to
$\Phi_{\intercal}:\mathcal{A}_{\intercal}\rightarrow\mathcal{B}_{\intercal}$
with respect to the $\lambda$-pairing of $\mathcal{A}$ with $\mathcal{A}%
_{\intercal}$ $\mu$-pairing (\ref{mupairing}):
\begin{equation}
\left\langle \Phi\left(  B\right)  ,\rho\right\rangle _{\lambda}=\left\langle
B,\Phi_{\intercal}\left(  \rho\right)  \right\rangle _{\mu}\;\;\;\forall
\rho\in\mathcal{A}_{\intercal},\;B\in\mathcal{B}. \label{duality}%
\end{equation}
In particular, if $\Phi_{\intercal}\left(  \rho\right)  =F_{\intercal}\rho
F_{\intercal}^{\dagger}$, then $\Phi\left(  B\right)  =F^{\dagger}BF$, where
$F$ is the transposed operator to $F_{\intercal}=\widetilde{F}$.

Recall that a normal map $\Phi:\mathcal{B}\rightarrow\mathcal{A}$ is CP iff
the map $\mathrm{I}_{0}\otimes\Phi_{\intercal}$ is positive on $\mathcal{B}%
_{\intercal}\left(  \ell^{2}\right)  \otimes\mathcal{A}_{\intercal}$ into
$\mathcal{B}_{\intercal}\left(  \ell^{2}\right)  \otimes\mathcal{B}%
_{\intercal}$, where $\mathrm{I}_{0}=\mathrm{id}$ is the identity map on
$\mathcal{B}_{\intercal}\left(  \ell^{2}\right)  $. Moreover, at least in the
case of the simple algebra $\mathcal{B=B}\left(  \mathcal{H}\right)  $, $\Phi$
is normal on $\mathcal{B}$ iff it is weak continuation of the CP map
$\Phi|\mathcal{B}_{0}\left(  \mathfrak{h}\right)  $. This means that $\Phi$
like a normal state is completely determined on finite rank operators
$B=\sum_{ik}|i\rangle b_{ik}\langle k|$ by the operator entries $\Phi_{\mu
}^{ik}=\Phi\left(  |i\rangle\langle k|\right)  \in\mathcal{A}$ of the
Hermitian-positive matrix $\Phi_{\mu}=\left[  \Phi_{\mu}^{ik}\right]  $, and
thus has the form $\Phi\left(  B\right)  =\left\langle B,\Phi_{\mu
}\right\rangle _{\mu}$ in terms of the $\left(  \mathcal{B},\mathcal{B}%
_{\intercal}\right)  _{\mu}$ pairing%
\[
\left\langle B,\Phi_{\mu}\right\rangle _{\mu}=\sum_{ik}b_{ik}\Phi_{\mu}%
^{ik},\;\;\;b_{ik}=\left\langle i|B|k\right\rangle .
\]
of this $\mathcal{B}$ with $\mathcal{B}_{\intercal}=\mathcal{B}_{\intercal
}\left(  \mathfrak{h}\right)  $ with respect to $\mu=\mathrm{tr}%
_{\mathfrak{h}}=\tau$. Here $\Phi_{\mu}=\sum|i\rangle\Phi_{\mu}^{ik}\langle
k|\geq0$ is a kernel on $\mathfrak{g}\otimes\mathfrak{h}$ given in an
orthogonal basis of real units $\langle i|\in\mathfrak{h}$ and $|i\rangle
=\langle i|^{\dagger}$ by the Hermit-positive $\mathcal{A}$-valued matrix
$\left[  \Phi_{\mu}^{ik}\right]  $ having the unit partial trace%
\[
\mu\left(  \Phi_{\mu}\right)  :=\sum_{i}\Phi_{\mu}^{ii}\equiv\left\langle
I_{\mathfrak{h}},\Phi_{\mu}\right\rangle _{\mu}=I_{\mathfrak{g}}%
\]
if $\Phi\left(  I_{\mathfrak{h}}\right)  =I_{\mathfrak{g}}$. The kernel
$\Phi_{\mu}$, called the \emph{density} of a quantum operation $\Phi
:\mathcal{B}\rightarrow\mathcal{A}$ with respect to the trace $\bar{\mu
}\left(  B\right)  =\mathrm{tr}_{\mathfrak{h}}B$ on $\mathcal{B}$, defines the
density operators of output states $\sigma=\rho\circ\Phi$ by the partial
tracing
\begin{equation}
\sigma=\tau\left[  \left(  \bar{\rho}\otimes I_{\mathfrak{h}}\right)
\Phi_{\mu}\right]  \equiv\left\langle \Phi_{\mu},\rho\right\rangle _{\lambda}.
\label{ptrace}%
\end{equation}
in terms of the $\left(  \mathcal{A},\mathcal{A}_{\intercal}\right)  $ pairing
with respect to the reference trace $\lambda=\bar{\tau}$ on $\overline
{\mathcal{A}}$. Note that the usual trace-form of the pairing defines the
covariant form $\bar{\sigma}=\tau\left(  \Phi_{\mu}^{\prime}\left(  \bar{\rho
}\otimes I_{\mathfrak{h}}\right)  \right)  $ of the output matrix $\sigma$ by
the partial transposition $\Phi_{\mu}^{\prime}=\left[  \Phi_{\mu}^{ki}\right]
$ which may be not Hermitian-positive for the positive $\Phi_{\mu}$. This
indicates that covariant trace-pairing is not natural for describing quantum
channelling and explains our preference to use the tensor form of the pairings
corresponding to the contravariant form of the trace-densities. Otherwise
there would not be one-to-one correspondence between complete positivity of
operations and Hermitian positivity of their densities as kernels.

In general the density $\Phi_{\mu}$ is an unbounded Hermitian-positive
operator, or even a generalized one, defined only as a kernel of a positive
Hermitian form in $\mathfrak{g}\otimes\mathfrak{h}$. Nevertheless we will
still use the notations of the partial tracings%
\[
\Phi_{\intercal}\left(  \rho\right)  =\left\langle \Phi_{\mu},\rho
\right\rangle _{\lambda},\;\;\Phi\left(  B\right)  =\left\langle B,\Phi_{\mu
}\right\rangle _{\mu}%
\]
and write $\mu\left(  \Phi_{\mu}\right)  =\left\langle I_{\mathfrak{h}}%
,\Phi_{\mu}\right\rangle _{\mu}$, defining the channel densities by the
Hermitian positivity and the normalization conditions%
\[
\Phi_{\mu}\geq0,\;\mu\left(  \Phi_{\mu}\right)  =I_{\mathfrak{g}}.
\]
condition $\Phi\left(  I_{\mathfrak{h}}\right)  =I_{\mathfrak{g}}$. In
particular, the density $\Phi_{\mu}$ of the operation $\Phi\left(  B\right)
=F^{\dagger}BF$ with respect to the standard trace $\mu=\mathrm{tr}%
_{\mathfrak{h}}$ is described as the kernel $\Phi_{\mu}=|F)(F|$, where the
generalized bra-vector $(F|$ is well-defined on the finite linear combinations
of the products $\xi\otimes\eta\in\mathfrak{g}\otimes\mathfrak{h}$ by the
partial transposition%
\begin{equation}
(F|\left(  |\xi\rangle\otimes|\eta\rangle\right)  =\eta F|\xi\rangle
\;\;\;\;\forall\xi\in\mathfrak{g},\eta\in\mathfrak{h}. \label{genbra}%
\end{equation}
It cannot be defined as an element of the Hilbert space $\mathcal{H}%
=\mathfrak{g}\otimes\mathfrak{h}$ for any isometry, a unitary operator $F=U$
say, if the space $\mathfrak{g}$ has infinite dimensionality $\dim
\mathfrak{g}=\infty$.

Note that for the arbitrary output algebra $\mathcal{B}$ the quantum channel
density as a positive self-adjoint, possibly unbounded operator (kernel)
affiliated to $\mathcal{A}\otimes\overline{\mathcal{B}}$ was first introduced
in \cite{BeSt86} with respect to operator-valued weights $\phi$ as CP maps
into $\mathcal{A}$ densely defined in $\mathcal{B}$, which correspond in our
case to $\phi\left(  B\right)  =$ $\nu\left(  B\right)  I_{\mathfrak{g}}$.
However, in order to avoid technicalities one can consider here only the
channels with the bounded densities $\Phi_{\mu}$ with respect to the trace
$\mu=\bar{\nu}$.

\section{An explicit formula for operational CB distance}

As a measure of difference between two quantum channels $\Phi$ and
$\Psi:\mathcal{B}\rightarrow\mathcal{A}$ one can adopt the usual boundedness
norm\emph{ }distance\emph{\ }$D_{\mathrm{b}}(\Phi,\Psi):=\Vert\Phi-\Psi
\Vert_{\mathrm{b}}$ (or B-distance for short), defined in terms of the density
operators $\Phi_{\mu}$ and $\Psi_{\mu}$ as\emph{\ }%
\[
D_{\mathrm{b}}(\Phi,\Psi)=\sup_{\rho\in\mathcal{S}_{\lambda}}\mu\left\vert
\Phi_{\intercal}\left(  \rho\right)  -\Psi_{\intercal}\left(  \rho\right)
\right\vert =\sup_{\rho\in\mathcal{S}_{\lambda}}\mu\left(  \left\vert
\tau\left[  \Delta_{\mu}\left(  \rho\right)  \right]  \right\vert \right)  ,
\]
where $\tau\left[  \Delta_{\mu}\left(  \rho\right)  \right]  $ is dual action
$\Delta_{\intercal}\left(  \rho\right)  =\left\langle \Delta_{\mu}%
,\rho\right\rangle _{\lambda}$ of $\Delta=\Phi-\Psi$ which us equal to the
partial tracing (\ref{ptrace}) of the operator%
\begin{equation}
\Delta_{\mu}\left(  \rho\right)  :=\left(  \tilde{\rho}^{1/2}\otimes
I_{\mathfrak{h}}\right)  \Delta_{\mu}\left(  \tilde{\rho}^{1/2}\otimes
I_{\mathfrak{h}}\right)  \in\mathcal{S}_{\tau\otimes\mu}\left(  \Phi
_{\intercal}\left(  \rho\right)  \right)  -\mathcal{S}_{\tau\otimes\mu}\left(
\Psi_{\intercal}\left(  \rho\right)  \right)  \subset\left(  \overline
{\mathcal{A}}\otimes\mathcal{B}\right)  _{\intercal} \label{stent}%
\end{equation}
in terms of the $\mu$-density operator $\Delta_{\mu}=\Phi_{\mu}-\Psi_{\mu}$.
Here $\mathcal{S}_{\tau\otimes\mu}\left(  \sigma\right)  $ denotes the convex
set of normal state density operators $\omega\in\left(  \overline{\mathcal{A}%
}\otimes\mathcal{B}\right)  _{\intercal}$ with respect to the product trace
$\tau\otimes\mu$ in $\mathcal{A}\otimes\overline{\mathcal{B}}$, having fixed
partial trace $\tau\left(  \omega\right)  =\sigma$.

However it is more appropriate to use the larger \emph{CB-distance
}$D_{\mathrm{cb}}(\Phi,\Psi):=\Vert\Delta\Vert_{\mathrm{cb}}$ since the
difference $\Delta=\Phi-\Psi$ of two CP maps is not just bounded, but it is
also completely bounded, $\Vert\Delta\Vert_{\mathrm{cb}}<\infty$, where
$\Vert\cdot\Vert_{\mathrm{cb}}\geq\Vert\cdot\Vert_{\mathrm{b}}$ is the
so-called norm of complete boundedness \cite{Pau03} (or CB-norm for short).
Instead of maximizing over normal input states $\rho\in\mathcal{S}_{\lambda}$
one should maximize over normal \emph{input entanglements} with any probe
quantum system $\mathbb{A}$, or at least with the standard one, described by
the "matrix algebra" $\mathbb{A}=\mathcal{B}\left(  \mathfrak{k}\right)
=\overline{\mathbb{A}}$ on the Hilbert space $\mathfrak{k}=\ell^{2}$ of all
square-summable sequences indexed by $\mathbb{N}$. The normal entanglements
are usually given by the densities $\hat{\rho}\in\mathcal{S}_{\hat{\tau}}$ of
the normal states $\hat{\rho}$ on the algebra $\mathbb{A}\otimes\mathcal{A}$
with respect to the standard trace $\hat{\tau}=\tau_{\mathfrak{k}}\otimes\tau$.

This maximization can be written as the supremum%
\[
D_{\mathrm{cb}}(\Phi,\Psi)=\sup_{\rho\in\mathcal{S}_{\lambda}}D_{\mathrm{cb}%
}^{\rho}\left(  \Phi,\Psi\right)
\]
over all input states of the \emph{conditional} CB semidistance as the maximal
trace distance
\begin{equation}
D_{\mathrm{cb}}^{\rho}\left(  \Phi,\Psi\right)  =\sup_{\hat{\rho}%
\in\mathcal{S}_{\hat{\tau}}}\left\{  \hat{\tau}\left\vert \check{\Phi
}_{\intercal}\left(  \hat{\rho}\right)  -\check{\Psi}_{\intercal}\left(
\hat{\rho}\right)  \right\vert :\tau_{\mathfrak{k}}\left(  \hat{\rho}\right)
=\rho\right\}  , \label{cbd}%
\end{equation}
of the entangled output states $\hat{\varphi}=\hat{\rho}\circ\check{\Phi}$ and
$\hat{\sigma}=\hat{\rho}\circ\check{\Psi}$ on $\mathbb{A}\otimes\mathcal{B}$
with respect to the input entangled states $\hat{\rho}$ having fixed margin
$\rho$ on $\mathcal{A}$. Here $\check{\Phi}_{\intercal}=\mathrm{Id}\otimes
\Phi_{\intercal}$ and $\check{\Psi}_{\intercal}=\mathrm{Id}\otimes
\Psi_{\intercal}$ are left ampliations $\mathbb{A}_{\intercal}\mathbb{\otimes
}\mathcal{A}_{\intercal}\rightarrow\mathbb{A}_{\intercal}\mathbb{\otimes
}\mathcal{B}_{\intercal}$ of $\Phi_{\intercal}$ and $\Psi_{\intercal}$ on the
predual space $\mathbb{A}_{\intercal}=\mathcal{B}_{\intercal}\left(
\mathfrak{k}\right)  $ of trace class matrices $\left[  \lambda\left(
\hat{\rho}^{ik}\right)  \right]  $ in the natural basis of $\ell^{2}$ and
$\mathcal{A}\subseteq\mathcal{B}\left(  \mathfrak{g}\right)  $, with
$\hat{\tau}=\tau_{\mathfrak{k}}\otimes\mu$. They are applied as $\check
{\Delta}_{\intercal}\left(  \hat{\rho}\right)  =\left[  \Delta_{\intercal
}\left(  \hat{\rho}^{ik}\right)  \right]  $ on the convex set $\mathcal{S}%
_{\hat{\tau}}\left(  \rho\right)  $ of the predual space$\ \left(
\mathbb{A\otimes}\mathcal{A}\right)  _{\intercal}=\mathbb{A}_{\intercal
}\mathbb{\otimes}\mathcal{A}_{\intercal}$ of all $\mathcal{A}_{\intercal}%
$-valued density matrices $\hat{\rho}=\left[  \hat{\rho}^{ik}\right]  $,
$\hat{\rho}^{ik}\in\mathcal{A}_{\intercal}$ with respect to the trace
$\hat{\tau}$ on $\overline{\mathbb{A}}\mathbb{\otimes}\overline{\mathcal{A}%
}=\mathbb{A\otimes}\overline{\mathcal{A}}$, having the partial trace
$\tau_{\mathfrak{k}}\left(  \hat{\rho}\right)  :=\sum_{i}\hat{\rho}^{ii}=\rho$.

Taking into account that any $\hat{\rho}\in\mathcal{S}_{\hat{\tau}}\left(
\rho\right)  $ can be written in blocks as $\left[  \rho^{1/2}\Pi^{ik}%
\rho^{1/2}\right]  $ with $\Pi^{ik}\in\overline{\mathcal{A}}$ defining a
normal unital CP map $\Pi\left(  \bar{A}\right)  =\bar{a}_{ik}\Pi^{ik}$ on the
matrix algebra $\mathbb{A}$ into $\overline{\mathcal{A}}$, we can write
\[
\Delta_{\intercal}\left(  \hat{\rho}^{ik}\right)  =\left\langle \Delta_{\mu
},\rho^{1/2}\Pi^{ik}\rho^{1/2}\right\rangle _{\lambda}=\left\langle \Pi
^{ik},\Delta_{\mu}\left(  \rho\right)  \right\rangle _{\tau}\equiv\hat{\Pi
}_{\intercal}^{ik}\left[  \Delta_{\mu}\left(  \rho\right)  \right]
\;\;\ \forall\hat{\rho}\in\mathcal{S}_{\hat{\tau}}\left(  \rho\right)  ,
\]
where $\hat{\Pi}=\Pi\otimes\mathrm{Id}$ is the right ampliation $\mathbb{A}%
\otimes\mathcal{B}\rightarrow\overline{\mathcal{A}}\otimes\mathcal{B}$ of the
normal unital CP map $\Pi$ given by the $\overline{\mathcal{A}}$-valued
density matrix $\left[  \Pi^{ik}\right]  $ with respect to $\tau
_{\mathfrak{k}}$. Thus the supremum in (\ref{cbd}) can be expressed in the
form of the supremum over all normal unital CP maps $\Pi$ as%
\begin{equation}
D_{\mathrm{cb}}^{\rho}(\Phi,\Psi)=\sup_{\Pi}\hat{\tau}\left\vert \hat{\Pi
}_{\intercal}\left[  \Delta_{\mu}\left(  \rho\right)  \right]  \right\vert
\leq\left(  \tau\otimes\mu\right)  \left\vert \Delta_{\mu}\left(  \rho\right)
\right\vert , \label{incbd}%
\end{equation}
where we used the obvious inequality $\left\Vert \hat{\Pi}_{\intercal}\left(
\omega\right)  \right\Vert _{\intercal}\leq\left\Vert \omega\right\Vert
_{\intercal}:=\left(  \tau\otimes\mu\right)  \left\vert \omega\right\vert $
valid for any linear trace preserving CP map $\Pi_{\intercal}$ on
$\overline{\mathcal{A}}_{\intercal}$ into $\mathbb{A}_{\intercal}%
=\mathcal{B}_{\intercal}\left(  \mathfrak{k}\right)  $ and $\omega\in\left(
\overline{\mathcal{A}}\otimes\mathcal{B}\right)  _{\intercal}$.

The inequality in (\ref{incbd}) is actually the equality%
\begin{equation}
D_{\mathrm{cb}}^{\rho}(\Phi,\Psi)=\left(  \tau\otimes\mu\right)  \left\vert
\left(  \tilde{\rho}^{1/2}\otimes I_{\mathfrak{h}}\right)  \left(  \Phi_{\mu
}-\Psi_{\mu}\right)  \left(  \tilde{\rho}^{1/2}\otimes I_{\mathfrak{h}%
}\right)  \right\vert , \label{cbformula}%
\end{equation}
in the case of separable $\mathcal{A}$ as it can easily be seen for the simple
algebras $\mathcal{A}=\mathcal{B}\left(  \mathfrak{g}\right)  $ when the
supremum is obviously achieved at $\Pi=\mathrm{Id}$ on the opposite input
algebra $\mathbb{A}=\overline{\mathcal{A}}$ coinciding with $\mathcal{B}%
\left(  \mathfrak{k}\right)  $ in a representation $\mathfrak{g}=\mathfrak{k}$.

In order to obtain the same formula for an arbitrary semifinite algebra
$\mathcal{A}$, the supremum in (\ref{cbd}) over $\hat{\rho}\in\mathcal{S}%
_{\hat{\tau}}\left(  \rho\right)  $ should also be extended to the nonstandard
$\mathbb{A}$. This can be seen as optimization of the operational distance via
all \emph{quantum encodings} \cite{BeOh02} as CP mappings%
\[
\pi\left(  \bar{A}\right)  =\rho^{1/2}\Pi\left(  \bar{A}\right)  \rho
^{1/2}=\bar{a}_{ik}\hat{\rho}^{ik},\;\;\;\bar{A}=\left[  \bar{a}_{ik}\right]
\]
of not only standard algebra $\mathbb{A}=\mathcal{B}\left(  \mathfrak{k}%
\right)  $ into $\mathcal{A}_{\intercal}$ but also less simple algebras,
including $\mathbb{A}=\overline{\mathcal{A}}$. Quantum encodings were defined
in \cite{BeOh02} by normal CP maps $\pi:\mathbb{A}\rightarrow\overline
{\mathcal{A}}$ on any semifinite "quantum alphabet" algebra $\mathbb{A}$ with
fixed normalization $\pi\left(  I_{\mathfrak{k}}\right)  =\rho$. They describe
\emph{normal couplings} of the corresponding state $\rho$ on $\mathcal{A}$
with the normal states on $\mathbb{A}$, having the densities $\mathrm{P}%
=\Pi_{\intercal}\left(  \bar{\rho}\right)  =\pi_{\intercal}\left(
I_{\mathfrak{g}}\right)  $ with respect to a trace $\tau$ on $\mathbb{A}$, by%
\[
\hat{\rho}=\left(  I_{\mathfrak{k}}\otimes\rho^{1/2}\right)  \Pi_{\bar{\tau}%
}\left(  I_{\mathfrak{k}}\otimes\rho^{1/2}\right)  \equiv\Pi_{\bar{\tau}%
}\left(  \rho\right)  ,
\]
where $\Pi_{\bar{\tau}}\in\mathbb{A}_{\intercal}\otimes\overline{\mathcal{A}}$
is the density operator of the unital CP map $\Pi$. The maximal distance over
all such encodings is obviously achieved on the \emph{standard coupling}
$\pi^{\ast}\left(  \bar{A}\right)  =\rho^{1/2}\bar{A}\rho^{1/2}$ as in the
case of the simple algebra $\mathcal{A}=\mathcal{B}\left(  \mathfrak{g}%
\right)  $, but the optimal "encoding alphabet" system algebra $\mathbb{A}%
=\overline{\mathcal{A}}$ does not coincide with $\mathcal{A}$ if
$\mathcal{A}\neq\overline{\mathcal{A}}$ but be antiisomorphic (opposite) to
$\mathcal{A}$. The standard entanglement defines the \emph{optimal compound
state}%
\[
\left\langle \bar{A}\otimes A,\omega\right\rangle _{\tau\otimes\bar{\tau}%
}=\bar{\tau}\left(  \upsilon\bar{A}\upsilon\widetilde{A}\right)  =\left(
\bar{\tau}\otimes\tau\right)  \left(  \bar{\omega}\left(  \bar{A}\otimes
A\right)  \right)  ,\;A\in\mathcal{A},\bar{A}\in\overline{\mathcal{A}},
\]
where $\upsilon=\rho^{1/2}$ and $\omega:=|\upsilon)(\upsilon|\equiv
\omega\left(  \rho\right)  $ is the optimal density operator $\hat{\rho}%
^{\ast}$ of this standard entangled state with respect to $\hat{\tau}%
=\tau\otimes\lambda$ and the partial trace $\tau\left(  \omega\right)  =\rho$.

Thus the formula (\ref{cbformula}) gives an expression for the\emph{
CB-distance} as the supremum
\begin{equation}
D_{\mathrm{cb}}\left(  \Phi,\Psi\right)  =\sup_{\rho\in\mathcal{S}_{\lambda}%
}\left(  \tau\otimes\mu\right)  \left\vert \left(  \check{\Phi}_{\intercal
}-\check{\Psi}_{\intercal}\right)  \left[  \omega\left(  \rho\right)  \right]
\right\vert =\sup_{\rho\in\mathcal{S}_{\lambda}}\left(  \tau\otimes\mu\right)
\left\vert \Delta_{\mu}\left(  \rho\right)  \right\vert \label{gencb}%
\end{equation}
of the conditional CB semidistance in terms of the standard input entangled
states $\omega\left(  \rho\right)  \in\left(  \overline{\mathcal{A}}%
\otimes\mathcal{A}\right)  _{\intercal}$. It is maximal trace-distance of the
optimally entangled states on $\overline{\mathcal{A}}\otimes\mathcal{B}$
described by the densities $\hat{\phi}=\check{\Phi}_{\intercal}\left(
\omega\right)  $ and $\hat{\sigma}=\check{\Psi}_{\intercal}\left(
\omega\right)  $ with the same partial trace $\mathrm{P}=\lambda\left(
\omega\right)  =\bar{\rho}$.

One can also show that the CB-distance is majorized by the natural
density-operator distance (\textit{complete distance}) $D_{\mathrm{c}}\left(
\Phi,\Psi\right)  \geq D_{\mathrm{cb}}\left(  \Phi,\Psi\right)  $ having the
particularly simple form%
\begin{equation}
D_{\mathrm{c}}\left(  \Phi,\Psi\right)  =\sup_{\rho\in\mathcal{S}_{\lambda}%
}\left\langle \mu\left(  \left\vert \Phi_{\mu}-\Psi_{\mu}\right\vert \right)
,\rho\right\rangle _{\lambda}. \label{Dist}%
\end{equation}
(Obviously $D_{\mathrm{b}}\left(  \Phi,\Psi\right)  \leq D_{\mathrm{c}}\left(
\Phi,\Psi\right)  $, but it is not so obvious that $D_{\mathrm{cb}}\left(
\Phi,\Psi\right)  \leq D_{\mathrm{c}}\left(  \Phi,\Psi\right)  $.) However
$D_{\mathrm{c}}$ is not equivalent to the CB-distance, that is closeness of
$\Psi_{n}$ to $\Phi$ in the sense $D_{\mathrm{cb}}\left(  \Phi,\Psi
_{n}\right)  \searrow0$ does not guarantee the closeness with respect to
$D_{\mathrm{c}}\left(  \Phi,\Psi_{n}\right)  \geq D_{\mathrm{cb}}\left(
\Phi,\Psi_{n}\right)  $, and it is difficult to give an operational meaning of
the optimality criterion defined by this distance. This is why an operational
fidelity distance is even more desirable for quantum channels than for states.

\section{Helinger distance and relative channel fidelity}

One can define a Helinger like operational square-distance%
\begin{equation}
d\left(  \Phi,\Psi\right)  ^{2}:=\sup_{\rho\in\mathcal{S}_{\lambda}}\mu\left(
\Phi_{\intercal}\left(  \rho\right)  +\Psi_{\intercal}\left(  \rho\right)
-2\sqrt{\Phi_{\intercal}\left(  \rho\right)  \Psi_{\intercal}\left(
\rho\right)  }\right)  \label{qfiddist}%
\end{equation}
between two quantum operations $\Phi$ and $\Psi$ as the Bures square-distance
$d\left(  \varphi,\sigma\right)  ^{2}$ for the output states $\varphi
=\rho\circ\Phi$, $\sigma=\rho\circ\Psi$ in terms of their output $\mu
$-densities $\sigma=\Psi_{\intercal}\left(  \rho\right)  $,%
\[
\Psi_{\intercal}\left(  \rho\right)  =\tau\left[  \left(  \tilde{\rho}\otimes
I_{\mathfrak{h}}\right)  \Psi_{\mu}\right]  =\tau\left[  \Psi_{\mu}\left(
\rho\right)  \right]
\]
and similar for $\phi=\Phi_{\intercal}\left(  \rho\right)  $ in the notation
of the previous section, maximized over the input states $\rho$. In the case
of the channels described by the normal unital CP maps $\Phi$ and $\Psi$ this
can be expressed as $d\left(  \Phi,\Psi\right)  ^{2}/2=1-f\left(  \Phi
,\Psi\right)  $ in terms of the the minimal fidelity
\[
f\left(  \Phi,\Psi\right)  =\inf_{\rho\in\mathcal{S}_{\lambda}}\mu\left(
\sqrt{\tau\left[  \left(  \tilde{\rho}\otimes I_{\mathfrak{h}}\right)
\Phi_{\mu}\right]  \tau\left[  \left(  \tilde{\rho}\otimes I_{\mathfrak{h}%
}\right)  \Psi_{\mu}\right]  }\right)
\]
of the output states over all inputs $\rho\in\mathcal{S}_{\lambda}$ with
respect to $\lambda=\bar{\tau}$. As for any two output states $\phi,\sigma$ on
$\mathcal{B}$ the following equivalence inequality obviously holds for this
Helinger (H-) distance:%
\[
d\left(  \Phi,\Psi\right)  ^{2}\leq D_{\mathrm{b}}\left(  \Phi,\Psi\right)
\leq2d\left(  \Phi,\Psi\right)  .
\]
However there is no such equivalence inequality between this fidelity distance
and the CB distance (\ref{cbd}), and it is difficult to give a definition of
this fidelity without reference to the input states $\rho$.

Since the map $\Phi$ and $\Psi$ are not just positive but CP, it is more
appropriate to define the \emph{complete} fidelity distance of the operations
$\Phi$ and $\Psi$ as the supremum%
\[
d_{\mathrm{c}}\left(  \Phi,\Psi\right)  =\sup_{\rho\in S_{\lambda}%
}d_{\mathrm{c}}^{\rho}\left(  \Phi,\Psi\right)
\]
of the maximal Bures distance%
\[
d_{\mathrm{c}}^{\rho}\left(  \Phi,\Psi\right)  =\sup_{\hat{\rho}\in
\mathcal{S}_{\hat{\tau}}}\left\{  d\left(  \hat{\rho}\circ\check{\Phi}%
,\hat{\rho}\circ\check{\Phi}\right)  :\tau_{\mathfrak{k}}\left(  \hat{\rho
}\right)  =\rho\right\}
\]
for the input-output entangled states $\hat{\varphi}=\hat{\rho}\circ
\check{\Phi}$ and $\hat{\sigma}=\hat{\rho}\circ\check{\Phi}$\ over the
densities $\hat{\rho}\in\left(  \mathbb{A}\otimes\mathcal{A}\right)
_{\intercal}$ describing quantum encodings of a fixed input state $\rho$,
similar to (\ref{cbd}). Since%
\[
\sup_{\hat{\rho}=\mathcal{S}_{\hat{\tau}}\left(  \rho\right)  }d\left(
\hat{\rho}\circ\check{\Phi},\hat{\rho}\circ\check{\Phi}\right)  ^{2}%
=\mu\left(  \Phi_{\intercal}\left(  \rho\right)  +\Psi_{\intercal}\left(
\rho\right)  \right)  -2\inf_{\hat{\rho}\in\mathcal{S}_{\hat{\tau}}\left(
\rho\right)  }\hat{\tau}\left(  \sqrt{\check{\Phi}_{\intercal}\left(
\hat{\rho}\right)  \check{\Psi}_{\intercal}\left(  \hat{\rho}\right)
}\right)  ,
\]
in the case of the unital $\Phi,\Psi$ this can be written as $d_{\mathrm{c}%
}^{\rho}\left(  \Phi,\Psi\right)  ^{2}/2=1-f_{\mathrm{c}}^{\rho}\left(
\Phi,\Psi\right)  $ in terms of the \emph{complete relative quantum channel
fidelity}%
\begin{equation}
f_{\mathrm{c}}^{\rho}\left(  \Phi,\Psi\right)  =\inf_{\hat{\rho}\in
\mathcal{S}_{\hat{\tau}}}\left\{  \hat{\tau}\left(  \sqrt{\check{\Phi
}_{\intercal}\left(  \hat{\rho}\right)  \check{\Psi}_{\intercal}\left(
\hat{\rho}\right)  }\right)  :\tau_{\mathfrak{k}}\left(  \hat{\rho}\right)
=\rho\right\}  \label{crf}%
\end{equation}
conditioned by an input state $\rho$. Here as in (\ref{cbd}) the minimization
is given over quantum encodings%
\[
\hat{\rho}=\left(  I_{\mathfrak{k}}\otimes\rho^{1/2}\right)  \Pi_{\tau}\left(
I_{\mathfrak{k}}\otimes\rho^{1/2}\right)  \in\mathcal{S}_{\hat{\tau}}\left(
\rho\right)
\]
described by normal unital CP maps $\Pi:\mathbb{A}\rightarrow\overline
{\mathcal{A}}$ in terms of their densities $\Pi_{\tau}$ with respect to
$\tau=\tau_{\mathfrak{k}}$ for a fixed $\rho\in\mathcal{S}_{\lambda}$. Using
the monotonicity \cite{OhPe93, NiCh00}%
\[
f\left(  \hat{\Pi}_{\intercal}\left[  \Phi_{\mu}\left(  \rho\right)  \right]
,\hat{\Pi}_{\intercal}\left[  \Psi_{\mu}\left(  \rho\right)  \right]  \right)
\geq f\left(  \Phi_{\mu}\left(  \rho\right)  ,\Psi_{\mu}\left(  \rho\right)
\right)
\]
of the quantum state relative fidelity with respect to such $\Pi$, we obtain
the inequality%
\[
f_{\mathrm{c}}^{\rho}\left(  \Phi,\Psi\right)  =\inf_{\Pi}\hat{\tau}\left(
\sqrt{\hat{\Pi}_{\intercal}\left[  \Phi_{\mu}\left(  \rho\right)  \right]
\hat{\Pi}_{\intercal}\left[  \Psi_{\mu}\left(  \rho\right)  \right]  }\right)
\geq\left(  \tau\otimes\mu\right)  \left(  \sqrt{\Phi_{\mu}\left(
\rho\right)  \Psi_{\mu}\left(  \rho\right)  }\right)  .
\]
This inequality is obviously equality achieved at $\Pi=\mathrm{Id}$ in the
case the simple input algebra $\mathcal{A}=\mathcal{B}\left(  \mathfrak{g}%
\right)  $ on $\mathfrak{g}=\ell^{2}$, and this lower bound is also achieved
for any separable $\mathcal{A}$. Thus we arrive to the following formula%
\begin{equation}
f_{\mathrm{c}}^{\rho}\left(  \Phi,\Psi\right)  =\left(  \tau\otimes\mu\right)
\left(  \sqrt{\Phi_{\mu}\left(  \rho\right)  ^{1/2}\Psi_{\mu}\left(
\rho\right)  \Phi_{\mu}\left(  \rho\right)  ^{1/2}}\right)  \label{crfex}%
\end{equation}
for the complete relative fidelity $f_{\mathrm{c}}$ of quantum channels in
terms of their densities $\Phi_{\mu}$ and $\Psi_{\mu}$ with respect to the
trace $\mu$ and a given input state $\rho$. This defines the complete Helinger
operational half-square semidistance for each $\rho$:%
\begin{equation}
d_{\mathrm{c}}^{\rho}\left(  \Phi,\Psi\right)  ^{2}/2=1-\left(  \tau\otimes
\mu\right)  \left(  \left\vert \Phi_{\mu}^{1/2}\left(  \tilde{\rho}\otimes
I_{\mathfrak{g}}\right)  \Psi_{\mu}^{1/2}\right\vert \right)  ,
\label{cqqfiddist}%
\end{equation}
where we used an equivalent representation (\ref{finfid}) for the fidelity
(\ref{crfex}) which will be derived in the next Section. Since this is simply
relative fidelity for two optimal input-output entangled states described by
the densities $\hat{\phi}=\Phi_{\mu}\left(  \rho\right)  $ and $\hat{\sigma
}=\Psi_{\mu}\left(  \rho\right)  $, the following equivalence inequality
obviously holds for each $\rho$:%
\[
d_{\mathrm{c}}^{\rho}\left(  \Phi,\Psi\right)  ^{2}\leq D_{\mathrm{cb}}^{\rho
}\left(  \Phi,\Psi\right)  \leq2d_{\mathrm{c}}^{\rho}\left(  \Phi,\Psi\right)
.
\]

By allowing the arbitrary semifinite quantum alphabet algebras $\mathbb{A}$ we
can extend this formula to the complete channel fidelity as the infimum%
\begin{equation}
f_{\mathrm{c}}\left(  \Phi,\Psi\right)  =\inf_{\rho\in\mathcal{S}_{\lambda}%
}\left(  \tau\otimes\mu\right)  \left(  \left\vert \Phi_{\mu}^{1/2}\left(
\tilde{\rho}\otimes I_{\mathfrak{g}}\right)  \Psi_{\mu}^{1/2}\right\vert
\right)  \label{genqfid}%
\end{equation}
over all input states $\rho$ on any algebra $\mathcal{A}$. It defines the
complete operational Helinger (CH-) distance by $d_{\mathrm{c}}^{2}%
/2=1-f_{\mathrm{c}}$ as the supremum of (\ref{cqqfiddist}) over all standard
encodings $\omega$ such that $\check{\Phi}_{\intercal}\left(  \omega\right)
=\Phi_{\mu}\left(  \rho\right)  $ and $\check{\Psi}\left(  \omega\right)
=\Psi_{\mu}\left(  \rho\right)  $, corresponding to the\ optimal quantum
alphabet algebra $\mathbb{A}=\overline{\mathcal{A}}$. Applying the equivalence
inequality (\ref{eq:fvd}) to these states and taking then the supremum over
all $\rho\in\mathcal{S}_{\lambda}$ we obtain the equivalence inequality
\begin{equation}
d_{\mathrm{c}}\left(  \Phi,\Psi\right)  ^{2}\leq D_{\mathrm{cb}}\left(
\Phi,\Psi\right)  \leq2d_{\mathrm{c}}\left(  \Phi,\Psi\right)  \label{cfvcbd}%
\end{equation}
also for the complete fidelity distance and the CB-distance (\ref{gencb}).

Thus the complete relative fidelity (\ref{crf}) refines the CB-norm
inequalities $D_{\mathrm{b}}\leq D_{\mathrm{cb}}\leq D_{\mathrm{c}}$ by
providing the channel fidelity distance $d_{\mathrm{c}}$ which satisfies the
equivalence inequalities (\ref{cfvcbd}) due to the inequality $D_{\mathrm{cb}%
}\leq2\sqrt{1-f_{\mathrm{c}}^{2}}$.The complete channel fidelity
$f_{\mathrm{c}}$ has clear operational meaning as the minimal relative
fidelity of the compound states achieved over all input-output entanglements
via two quantum channels described by the densities $\Phi_{\mu},\Psi_{\mu}$
\cite{BeOh02}. In particular, as we show in the next section%
\begin{equation}
f_{\mathrm{c}}\left(  \Phi,\Psi\right)  =\inf_{\rho\in\mathcal{S}_{\lambda}%
}\left(  \sum_{j}\left\vert \tau\left(  \tilde{\rho}F_{j}^{\dagger}V\right)
\right\vert ^{2}\right)  ^{1/2} \label{cfidpure}%
\end{equation}
if $\Phi$ is given in Kraus form $\Phi\left(  B\right)  =\sum F_{j}^{\dagger
}BF_{j}$ and $\Psi$ is pure $\Psi\left(  B\right)  =V^{\dagger}BV$, given by
an isometry $V$.

\section{Operational CH-distance as a minimax problem}

Following analogy with quantum Bures distance as a variational problem
(\ref{statehdist}) we should define the operational\emph{ Bures distance} as
the square root of the natural quadratic distance between generalized Schmidt
decompositions $\Gamma,\Upsilon\in\mathcal{B}\left(  \mathfrak{f}%
\otimes\mathfrak{g}\right)  $ of their density operators $\Phi_{\mu}$ and
$\Psi_{\mu}$:
\begin{equation}
d_{\mathrm{c}}^{\ast}\left(  \Phi,\Psi\right)  ^{2}=\inf_{\Gamma,\Upsilon
}\left\{  \left\Vert \mu\left[  \left(  \Gamma-\Upsilon\right)  ^{\dagger
}\left(  \Gamma-\Upsilon\right)  \right]  \right\Vert :\Gamma^{\dagger}%
\Gamma=\Phi_{\mu},\Upsilon^{\dagger}\Upsilon=\Psi_{\mu}\right\}  .
\label{qqfiddist}%
\end{equation}
without explicit reference to the input state $\rho$. In the case of the
standard trace $\mu$ induced by $\mathrm{tr}_{\mathfrak{h}}$ this can be
written as the minimization of the Hilbert module square distance
\[
d_{\mathrm{c}}^{\ast}\left(  \Phi,\Psi\right)  ^{2}=\inf_{\boldsymbol{F,V}%
}\left\{  \left\Vert \left(  \boldsymbol{F}-\boldsymbol{V}\right)  ^{\dagger
}\left(  \boldsymbol{F}-\boldsymbol{V}\right)  \right\Vert :|\boldsymbol{F}%
)(\boldsymbol{F}|=\Phi_{\mu},|\boldsymbol{V})(\boldsymbol{V}|=\Psi_{\mu
}\right\}
\]
over equivalent Kraus \cite{Kra83} decompositions%
\begin{equation}
\Phi\left(  B\right)  =\sum_{j}F_{j}^{\dagger}BF_{j}\equiv\boldsymbol{F}%
^{\dagger}B\boldsymbol{F},\;\;\Psi\left(  B\right)  =\sum_{j}V_{j}^{\dagger
}BV_{j}\equiv\boldsymbol{V}^{\dagger}B\boldsymbol{V} \label{kraus}%
\end{equation}
corresponding to the purified Schmidt decomposition
\begin{equation}
\Phi_{\mu}=\sum|F_{j})(F_{j}|\equiv\Gamma^{\dagger}\Gamma,\;\Psi_{\mu}%
=\sum_{j}|V_{j})(V_{j}|\equiv\Upsilon^{\dagger}\Upsilon. \label{gsmidt}%
\end{equation}
Here $(\boldsymbol{F}|$ and $(\boldsymbol{V}|$ are the columns of
$(F_{j}|,(V_{j}|$ which are the components of%
\[
\Gamma=\sum_{j}|j\rangle(F_{j}|,\;\Upsilon=\sum_{j}|j\rangle(V_{j}|,
\]
in an orthonormal basis $\left\{  |j\rangle\right\}  $ of $\mathcal{H}$, say
$|j\rangle=|i\rangle\otimes|k\rangle\equiv|i,k\rangle$, defining the bounded
operators $F_{j},V_{j}:\mathfrak{g}\rightarrow\mathfrak{f}$ as acting on the
right of the bra-vectors $\langle k|\in\mathfrak{g}$:%
\[
\langle k|F_{j}|i\rangle=(F_{j}|\left(  |i\rangle\otimes|k\rangle\right)
=\langle j|\Gamma|i,k\rangle,\;\langle k|V_{j}|i\rangle=(V_{j}|\left(
|i\rangle\otimes|k\rangle\right)  =\langle j|\Gamma|i,k\rangle.
\]
over all input densities $\rho\in\mathcal{S}_{\lambda}$ with respect to
$\lambda=\bar{\tau}$.

Taking into account that $\left\Vert A^{\dagger}A\right\Vert =\sup_{\rho
\in\mathcal{S}_{\lambda}}\tau\left(  \tilde{\rho}A^{\dagger}A\right)  $, and
that the positive function%
\[
c\left(  \Gamma-\Upsilon;\rho\right)  =\left(  \tau\otimes\mu\right)  \left[
\left(  \Gamma-\Upsilon\right)  \left(  \tilde{\rho}\otimes I_{\mathfrak{g}%
}\right)  \left(  \Gamma-\Upsilon\right)  ^{\dagger}\right]
\]
given by the total trace $\tau\otimes\mu$ is convex with respect to
$\Gamma-\Upsilon$ and concave with respect to $\rho$, we can rewrite the
minimal distance (\ref{qqfiddist}) in the form%
\begin{equation}
d_{\mathrm{c}}^{\ast}\left(  \Phi,\Psi\right)  ^{2}=\inf_{\substack{\Gamma
:\Gamma^{\dagger}\Gamma=\Phi_{\mu}\\\Upsilon:\Upsilon^{\dagger}\Upsilon
=\Psi_{\mu}}}\sup_{\rho\in\mathcal{S}_{\lambda}}c\left(  \Gamma-\Upsilon
;\rho\right)  =\sup_{\rho\in\mathcal{S}_{\lambda}}\inf_{\substack{\Gamma
:\Gamma^{\dagger}\Gamma=\Phi_{\mu}\\\Upsilon:\Upsilon^{\dagger}\Upsilon
=\Psi_{\mu}}}c\left(  \Gamma-\Upsilon;\rho\right)  , \label{exchange}%
\end{equation}
where we exchanged the extrema since since $\mathcal{S}_{\lambda}$ is convex
and the function $c$ is actually a positive quadratic form with respect to
$\Gamma-\Upsilon$, certainly achieving its infimum. Thus, (\ref{qqfiddist})
can be represented in the form%
\[
d_{\mathrm{c}}^{\ast}\left(  \Phi,\Psi\right)  ^{2}=\sup_{\rho\in
\mathcal{S}_{\lambda}}\left\{  \left(  \tau\otimes\mu\right)  \left[  \left(
\Phi_{\mu}+\Psi_{\mu}\right)  \left(  \tilde{\rho}\otimes I_{\mathfrak{g}%
}\right)  \right]  -2f_{\mathrm{c}}^{\ast}\left(  \Phi,\Psi;\rho\right)
\right\}  ,
\]
where the first term achieves the value $2$ for the unital $\Phi$ and $\Psi$,
and%
\[
f_{\mathrm{c}}^{\ast}\left(  \Phi,\Psi;\rho\right)  =\sup_{\Gamma^{\dagger
}\Gamma=\Phi_{\mu},\Upsilon^{\dagger}\Upsilon=\Psi_{\mu}}\operatorname{Re}%
\left(  \tau\otimes\mu\right)  \left[  \left(  \tilde{\rho}\otimes
I_{\mathfrak{g}}\right)  \Gamma^{\dagger}\Upsilon\right]  .
\]

Now we will prove that this supremum is in fact the conditional relative
fidelity (\ref{crfex}), and the complete relative fidelity (\ref{genqfid}) for
quantum channels coincides with minimax
\begin{equation}
f_{\mathrm{c}}^{\ast}\left(  \Phi,\Psi\right)  =\inf_{\rho\in\mathcal{S}%
_{\lambda}}\sup_{\Gamma^{\dagger}\Gamma=\Phi_{\mu},\Upsilon^{\dagger}%
\Upsilon=\Psi_{\mu}}\operatorname{Re}\left(  \tau\otimes\mu\right)  \left[
\left(  \tilde{\rho}\otimes I_{\mathfrak{g}}\right)  \Gamma^{\dagger}%
\Upsilon\right]  \label{qcfid}%
\end{equation}
for the real part of the Hilbert-Schmidt scalar product $\left(  \tau
\otimes\mu\right)  \left(  X^{\dagger}Y\right)  $ of $X=\Gamma\left(
\tilde{\rho}\otimes I_{\mathfrak{g}}\right)  ^{1/2}$ and $Y=\Upsilon\left(
\tilde{\rho}\otimes I_{\mathfrak{g}}\right)  ^{1/2}$ with respect to the total
trace $\tau\otimes\mu$. Since the operators $X$ and $Y$ define the Schmidt
decompositions $\Phi_{\mu}\left(  \rho\right)  =X^{\dagger}X$ and $\Psi_{\mu
}\left(  \rho\right)  =Y^{\dagger}Y$ of the density operators $\hat{\Phi
}_{\intercal}\left[  \omega\left(  \rho\right)  \right]  $ and $\hat{\Psi
}_{\intercal}\left[  \omega\left(  \rho\right)  \right]  $ for the optimal
entangled states on the algebra $\mathbb{A}\otimes\mathcal{B}=\overline
{\mathcal{A}}\otimes\mathcal{B}$, the complete fidelity (\ref{genqfid}) is
simply the minimax relative fidelity for the input-output entangled\ states
obtained by two different transformations $\check{\Phi}_{\intercal}$ and
$\check{\Psi}_{\intercal}$ of the input standard entangled states on algebra
$\mathbb{A}\otimes\mathcal{A}=\overline{\mathcal{A}}\otimes\mathcal{A}$. It is
first maximized over all Schmidt decompositions corresponding to the Kraus
decompositions (\ref{kraus}) of $\Phi$ and $\Psi$, and then minimized with
respect to all density operators $\rho$ of the input states.

\begin{lemma}
Let $R,S\in\mathcal{A}\otimes\overline{\mathcal{B}}$ be positive bounded
operators such that they have finite trace $\tau\otimes\mu$. Then%
\begin{equation}
\sup_{X,Y}\left\{  \left(  \tau\otimes\mu\right)  \left(  X^{\dagger
}Y+Y^{\dagger}X\right)  :X^{\dagger}X=R,Y^{\dagger}Y=S\right\}  =2\left(
\tau\otimes\mu\right)  \left(  T\right)  , \label{lemsup}%
\end{equation}
where $T=\sqrt{XSX^{\dagger}}$. This supremum is achieved at any
$X\in\mathcal{A}\otimes\overline{\mathcal{B}}$ satisfying the condition
$X^{\dagger}X=R$, $X=R^{1/2}\equiv X_{\mathrm{o}}$ say, and at
$Y=Y_{\mathrm{o}}$ satisfying the equation $Y_{\mathrm{o}}X^{\dagger
}=T=XY_{\mathrm{o}}^{\dagger}$
\end{lemma}

\begin{proof}
First we observe, by applying the Schwarz inequality%
\begin{align*}
\left(  \tau\otimes\mu\right)  \left(  X^{\dagger}Y+Y^{\dagger}X\right)   &
\leq2\sqrt{\left(  \tau\otimes\mu\right)  \left(  X^{\dagger}X\right)  \left(
\tau\otimes\mu\right)  \left(  Y^{\dagger}Y\right)  }\\
&  =2\sqrt{\left(  \tau\otimes\mu\right)  \left(  R\right)  \left(
\tau\otimes\mu\right)  \left(  S\right)  }\leq\infty,
\end{align*}
that this supremum is finite. In order to find it one can use Lagrangian
multiplier method. Fixing $X$ satisfying $X^{\dagger}X=R$, $X_{\mathrm{o}%
}=R^{1/2}$ say, we can write the Lagrangian function as%
\[
\ell=\left(  \tau\otimes\mu\right)  \left(  X^{\dagger}Y+Y^{\dagger
}X-Y^{\dagger}YL\right)  ,
\]
where $L=L^{\dagger}\in\mathcal{A}\otimes\overline{\mathcal{B}}$ is the
Lagrangian multiplier corresponding to the Hermitian condition $S=Y^{\dagger
}Y=S^{\dagger}$. At the stationary point%
\[
\delta\ell=\left(  \tau\otimes\mu\right)  \left[  \left(  X^{\dagger
}-LY^{\dagger}\right)  \delta Y+\left(  X-YL\right)  \delta Y^{\dagger
}\right]  =0,
\]
so $Y=Y_{\mathrm{o}}$ must satisfy the equation $YL=X$ (the other equation
$LY^{\dagger}=X^{\dagger}$ is Hermitian adjoint, corresponding to $Y^{\dagger
}=Y_{\mathrm{o}}^{\dagger}$). Thus $Y_{\mathrm{o}}=XL^{-1}$, where $L^{-1}$
should be determined from $L^{-1}X^{\dagger}XL^{-1}=S$. Multiplying this from
the left by $X$ and from the right by $X^{\dagger}$ this gives $\left(
XL^{-1}X^{\dagger}\right)  ^{2}=XSX^{\dagger}$, or $XL^{-1}X^{\dagger}%
=\sqrt{XSX^{\dagger}}$. Thus, indeed $Y_{\mathrm{o}}X^{\dagger}=\sqrt
{XSX^{\dagger}}=XY_{\mathrm{o}}^{\dagger}$, and therefore%
\begin{equation}
\left(  \tau\otimes\mu\right)  \left(  Y_{\mathrm{o}}X^{\dagger}%
+XY_{\mathrm{o}}^{\dagger}\right)  =2\left(  \tau\otimes\mu\right)  \left(
\sqrt{XSX^{\dagger}}\right)  . \label{lemproofsup}%
\end{equation}
This extremal value is the maximal value because of convexity of the
maximizing function in (\ref{lemsup}). Note that due to $\sqrt{U^{\dagger
}XSX^{\dagger}U}=U^{\dagger}\sqrt{XSX^{\dagger}}U$ for any unitary $U$, the
supremum (\ref{lemsup}), coinciding with (\ref{lemproofsup}) does not depend
on the choice of $X$ satisfying $X^{\dagger}X=R$. Indeed, by virtue of polar
decomposition $X=UR^{1/2}$%
\[
2\left(  \tau\otimes\mu\right)  \left(  \sqrt{XSX^{\dagger}}\right)  =2\left(
\tau\otimes\mu\right)  \left(  U^{\dagger}\sqrt{XSX^{\dagger}}U\right)
=2\left(  \tau\otimes\mu\right)  \left(  \sqrt{R^{1/2}SR^{1/2}}\right)  .
\]
Rewriting the trace $\left(  \tau\otimes\mu\right)  \left(  T\right)  =\left(
\tau\otimes\mu\right)  \left(  Y_{\mathrm{o}}X^{\dagger}\right)  $ as $\left(
\tau\otimes\mu\right)  \left(  X^{\dagger}Y_{\mathrm{o}}\right)  $ with
\[
X^{\dagger}Y_{\mathrm{o}}=R^{1/2}\sqrt{R^{1/2}SR^{1/2}}R^{-1/2}\equiv\sqrt{RS}%
\]
corresponding to $X=R^{1/2}$, we obtain the optimal value in (\ref{lemsup}) in
either form $\left(  \tau\otimes\mu\right)  \left(  T\right)  =\left(
\tau\otimes\mu\right)  \left(  \sqrt{RS}\right)  $ or $\left(  \tau\otimes
\mu\right)  \left(  T\right)  =\left(  \tau\otimes\mu\right)  \left(
\sqrt{R^{1/2}SR^{1/2}}\right)  $.
\end{proof}

Note that choosing $X=R^{1/2}$ for $R=\Phi_{\mu}\left(  \rho\right)  $ and
$Y=R^{-1/2}\sqrt{RS}$ for $S=\Psi_{\mu}\left(  \rho\right)  $ gives the
formula (\ref{crfex}). This proves that the operational Bures distance
$d_{\mathrm{c}}^{\ast}$ and the operational Helinger distance $d_{\mathrm{c}}$
are the same: $d_{\mathrm{c}}^{\ast}=d_{\mathrm{c}}$. Another useful
representation gives the choice $X=\Phi_{\mu}^{1/2}\left(  \tilde{\rho}\otimes
I_{\mathfrak{g}}\right)  ^{1/2}$ corresponding to $\Gamma=\Phi_{\mu}^{1/2}$
gives the equivalent formula%
\begin{align}
f_{\mathrm{c}}^{\rho}\left(  \Phi,\Psi\right)   &  =\left(  \tau\otimes
\mu\right)  \left(  \left\vert \Phi_{\mu}^{1/2}\left(  \tilde{\rho}\otimes
I_{\mathfrak{g}}\right)  \Psi_{\mu}^{1/2}\right\vert \right)  : \label{finfid}%
\\
&  =\left(  \tau\otimes\mu\right)  \left(  \sqrt{\Phi_{\mu}^{1/2}\left(
\tilde{\rho}\otimes I_{\mathfrak{g}}\right)  \Psi_{\mu}\left(  \tilde{\rho
}\otimes I_{\mathfrak{g}}\right)  \Phi_{\mu}^{1/2}}\right)
\end{align}
for the complete relative fidelity (\ref{crfex}) conditioned by the input state.

\section{Example: Relative fidelity of a pure quantum channel}

Let us consider the case of pure quantum channel $\Psi\left(  B\right)
=V^{\dagger}BV$ given on $\mathcal{B}=\mathcal{B}\left(  \mathfrak{h}\right)
$ by an isometry $V$, $V^{\dagger}V=I_{\mathfrak{g}}$. In this case the input
states are mapped as $\Psi_{\intercal}\left(  \rho\right)  =V_{\intercal
}^{\dagger}\rho V_{\intercal}$, where $V_{\intercal}=\widetilde{V}$, the
Hilbert space transposition of $V$, and the density operator $\Psi_{\mu}$ with
respect to the standard trace $\mu\left(  B\right)  =\mathrm{tr}%
_{\mathfrak{h}}B$ is $\Psi_{\mu}=|V)(V|$, where the generalized bra-vector
$(V|$\ is defined in (\ref{genbra}), and $|V)=(V|^{\dagger}$ is its Hermitian
adjoint. Since
\[
(L|=(V|\left(  \tilde{\rho}\otimes I_{\mathfrak{g}}\right)  \Phi_{\mu}^{1/2}%
\]
is defined as well-defined as Hilbert space element,%
\[
(L|L)=\left(  V|\left(  \tilde{\rho}\otimes I_{\mathfrak{h}}\right)  \Phi
_{\mu}\left(  \tilde{\rho}\otimes I_{\mathfrak{h}}\right)  |V\right)  \leq1,
\]
we can use the formula $\sqrt{|L)(L|}=|L)\left(  L|L\right)  ^{-\frac{1}{2}%
}(L|$ in (\ref{finfid}). Thus we obtain in this simple case%
\begin{equation}
f_{\mathrm{c}}^{\rho}\left(  \Phi,\Psi\right)  =\left(  V|\left(  \tilde{\rho
}\otimes I_{\mathfrak{h}}\right)  \Phi_{\tau}\left(  \tilde{\rho}\otimes
I_{\mathfrak{h}}\right)  |V\right)  ^{1/2}, \label{finalfidpure}%
\end{equation}
as $\left(  \tau\otimes\mu\right)  \left(  |L)(L|\right)  =\left(  L|L\right)
$ is induced by standard trace $\operatorname{Tr}=\mathrm{tr}_{\mathfrak{g}%
\otimes\mathfrak{h}}$. Using the Kraus decomposition $\Phi\left(  B\right)
=\sum F_{j}^{\dagger}BF_{j}$, this can be written as
\[
f_{\mathrm{c}}^{\rho}\left(  \Phi,\Psi\right)  =\left(  \sum_{j}\left\vert
\left(  V|\left(  \tilde{\rho}\otimes I_{\mathfrak{h}}\right)  |F_{j}\right)
\right\vert ^{2}\right)  ^{\frac{1}{2}}=\left(  \sum_{j}\left\vert
\mathrm{tr}_{\mathfrak{g}}\left(  \tilde{\rho}F_{j}^{\dagger}V\right)
\right\vert ^{2}\right)  ^{\frac{1}{2}}%
\]
where we used the identity $\left(  V|\left(  \tilde{\rho}\otimes
I_{\mathfrak{h}}\right)  |F\right)  =\mathrm{tr}_{\mathfrak{g}}\left(
\tilde{\rho}F^{\dagger}V\right)  $. This defines the expression used in
(\ref{cfidpure}), and if $\Phi$ is also pure, $\Phi\left(  B\right)
=F^{\dagger}BF$, we obtain even simpler formula $f_{\mathrm{c}}^{\rho}\left(
\Phi,\Psi\right)  =\left\vert \mathrm{tr}_{\mathfrak{g}}\left(  \tilde{\rho
}F^{\dagger}V\right)  \right\vert $. Therefore, the CF-distance of such $\Phi$
and $\Psi$ is evaluated in $d_{\mathrm{c}}^{2}/2=1-f_{\mathrm{c}}^{\rho
}\left(  \Phi,\Psi\right)  $ by the infimum%
\[
f_{\mathrm{c}}\left(  \Phi,\Psi\right)  =\inf_{\rho\in\mathcal{S}_{\lambda}%
}\left(  \sum_{j}\left\vert \mathrm{tr}_{\mathfrak{g}}\left(  \tilde{\rho
}F_{j}^{\dagger}V\right)  \right\vert ^{2}\right)  ^{1/2}.
\]
which is minimal magnitude $\inf_{\rho}\left\vert \mathrm{tr}_{\mathfrak{g}%
}\left(  \tilde{\rho}F^{\dagger}V\right)  \right\vert $ of correlation between
$F$ and $V$ in the case of pure $\Phi$. Note that the CB distance cannot be so
explicitly evaluated in the case of one pure channel, and it does not have
such simple interpretation even when both channels are pure.

\end{document}